\documentclass[12pt,preprint]{aastex}

\begin{document}


\title{The Distribution of Ly$\alpha$-Emitting Galaxies at z=2.38: Paper 2, Spectroscopy}


\author{Paul J. Francis\altaffilmark{1}}
\affil{Research School of Astronomy and Astrophysics, the Australian
National University, Canberra 0200, Australia}
\email{pfrancis@mso.anu.edu.au}

\altaffiltext{1}{Joint Appointment with the Department of Physics, 
Faculty of Science, the Australian National University}

\author{Povilas Palunas}
\affil{McDonald Observatory, University of Texas, Austin, TX 78712}

\author{Harry I. Teplitz}
\affil{Spitzer Science Center, California Institute of
Technology, Mail Code 220-62, 770 South Wilson Avenue, Pasadena, CA 91125}

\author{Gerard M. Williger}
\affil{Dept. of Physics and Astronomy, Johns Hopkins University, 3701
San Martin Drive, Baltimore, MD 21218}

\and

\author{Bruce E. Woodgate}
\affil{NASA Goddard Space Flight Center, Code 681, Greenbelt, MD 20771}

\begin{abstract}

In Paper 1 of this series we identified an 80 co-moving Mpc filament of 
candidate Ly$\alpha$ emitting galaxies at redshift 2.38. In this paper we present 
spectroscopy of the 37 galaxy candidates. Our 
spectroscopy reached a surface brightness limit of 
$5.0 \times 10^{-17}{\rm erg\ cm}^{-2}{\rm s}^{-1}{\rm arcsec}^{-2}$.
Of the 14 candidates down to this limit, 12 were confirmed to be 
Ly$\alpha$ emitting galaxies at the filament redshift. We also obtained spectral 
confirmation for six of the lower surface brightness candidates, all of which 
also lay at the filament redshift.
In addition, we identify a foreground cluster of QSOs at $z=1.65$.

\end{abstract}


\section{Introduction}

A large fraction of galaxies today lie in filamentary structures, such as the
Great Wall \citep{gel89}, separated by voids. When did these structures form?
There are now some tentative observations suggesting that this topology was
already in place as early as redshift three 
\citep[eg.][]{cam99,mol01}.

We recently carried out a large scale survey to measure the topology of the
distribution of Ly$\alpha$ emitting galaxies at $z \sim 2.38$ 
\citep[][hereafter Paper 1]{pal04}. We searched for 
candidate Ly$\alpha$ emitting galaxies in an 80$\times$ 80 $\times$ 60 
co-moving Mpc region, using the now well established narrow-band 
selection technique \citep[eg.][]{hu96,ste00,kud00,ven02,rho03,ouc03,sta04}. The region was
imaged through a narrow-band (54\AA ) filter, centered at 4110\AA . This was 
sensitive to Ly$\alpha$ emitting galaxies at $2.36 < z < 2.40$. 37 candidate
galaxies were found.

These candidate $z \sim 2.38$ galaxies mostly lie in a filamentary structure,
over 80 co-moving Mpc in length. This filament is 
seemingly bracketed by voids on both sides. In Paper~1 we argue that this
structure is statistically significant, and that the size of the voids and
the filament is larger than would be expected from CDM (Cold Dark Matter)
simulations. 

We did not, however, have spectra of the candidate high
redshift galaxies. It was thus possible that some or all of them were actually
foreground galaxies: galaxies lying at $z \sim 0.1$ whose \ion{O}{2} 3727\AA\
line lies within the filter bandpass. Even if they were Ly$\alpha$ emitting
galaxies at $z \sim 2.38$, their observations only constrained the two
dimensional position of the galaxies and not their precise redshift. 

In this paper, we present confirmation spectroscopy of these galaxy candidates.
We demonstrate that candidates are indeed primarily $z \sim 2.38$ Ly$\alpha$ emitting
galaxies, and we constrain the three-dimensional shape of the claimed filament.

\section{Observations and Reduction\label{obsred}}

Our spectroscopy was carried out using the Two Degree Field (2dF)
multi-fiber
spectrograph on the Anglo-Australian Telescope \citep[AAT,][]{lew02}.
This spectrograph has 400 fibers, spread over a circular field of diameter 
two degrees, located at the prime focus of the AAT. 

Fibers were allocated to targets using the {\it configure} program
\citep[][]{lew02}. First priority in the fiber allocation was given to
the candidate z=2.38 galaxies identified by \citet{pal04} (the main sample). 
We were able to
allocate fibers to all but two of these (the two unobserved sources, those
at  21:40:33.1 -44:36:10.8 and 21:43:05.9 -44:27:21.0, J2000, lay too close 
to other candidate galaxies for fiber allocation). Second priority was 26
brighter point sources which also showed excess narrow-band flux: these were
potential z=2.38 QSOs. Three of these were too bright for us to observe, but 
we were able to allocate fibers to twenty of the remaining twenty-three (the 
point source sample).

The candidate list of \citet{pal04} only included sources with excess 
narrow-band
emission of equivalent width $> 125$\AA . This was done to minimize
contamination from foreground [\ion{O}{2}] emitting galaxies, but may also
have eliminated some Ly$\alpha$ emitting sources. To check this, we selected
a sample of 86 sources which showed excess emission in the narrow band, but 
not enough to meet this equivalent width threshold (the low equivalent width 
sample). We were able to allocate fibers to 70 of these sources.

211 UVX sources were selected, as possible foreground QSOs. We were able to
allocate fibers to 120 of these (the UVX sample). Four candidate 
color-selected high redshift
QSOs were also observed.

Fibers were allocated to 17 Chandra sources: these will be discussed by 
Williger et al (in preparation).

\subsection{Observations}

Observations were carried out on the nights of 2003 August 31 and September 1.
Conditions were poor: the first night was mostly cloudy, and the second night,
while clear, suffered from $ \sim 2.5\arcsec$ seeing. We obtained a total
integration time of 5400 sec on the first night and 32,400 sec on the second
night, though much of this was obtained at high airmass, or while the moon was
up. 

The first night's data
were reduced quickly, allowing us to identify several foreground galaxies
amongst the low equivalent width sample. The fiber configuration was 
changed for the second night by eliminating these sources, allowing us to
observe more of the various lower priority samples. This also means that the
higher priority sources were observed through different fibers on the different
nights, giving us a check against systematic errors.

We used the 600V gratings in both the 2dF spectrographs, centered at a
wavelength of 4982\AA . This gave a spectral resolution of 450 ${\rm km\
s}^{-1}$ and a wavelength range of 3890 -- 6080\AA . The data were reduced 
using the {\it 2dfdr} software \citep[][]{lew02}.

Spectra from each night were co-added separately, as a check on the reality of
any faint features seen. A weighted sum of the spectra from both nights was 
then used in the final analysis.

\subsection{Spectral Classification}

All spectra were classified interactively. For
sources showing a single narrow emission line at $\sim 4110$\AA , we
classified them either as $z\sim 2.38$ Ly$\alpha$ emitting galaxies or
$z \sim 0.1$ [\ion{O}{2}] emitting blue compact galaxies using the following criteria:

\begin{itemize}

\item If we see corresponding [\ion{O}{3}] and/or H$\beta$ emission lines,
we identify the source as an [\ion{O}{2}] emitting galaxy at $z \sim 0.1$.

\item If we see corresponding \ion{C}{4} emission line,
we identify the source as a Ly$\alpha$ emitting galaxy at $z \sim 2.38$.

\item If the $\sim 4110$\AA\ line has a velocity width 
$> 500 {\rm km\ s}^{-1}$, and no
other lines are seen, we classify the line as Ly$\alpha$ at $z \sim 2.38$, on
the basis that the [\ion{O}{2}] emission of a blue compact galaxy is unlikely 
to be this broad.

\item The remaining three sources (with single narrow lines) were classified 
as Ly$\alpha$ emitting galaxies at $z \sim 2.38$, on the basis that had 
they been at $z \sim 0.1$, we should have been seen [\ion{O}{3}] emission 
(our $z \sim 0.1$ galaxies have a median flux calibrated ratio of 
[\ion{O}{3}]/[\ion{O}{2}]= 1.7). One of these three
was confirmed by \citet{fra97} to lie at $z \sim 2.38$.

\end{itemize}

The detection of an emission-line is regarded as secure if it is seen at
greater than $5 \sigma$ confidence. If seen at between 3 and 5 $\sigma$
confidence, it is regarded as marginal.

\section{Results\label{results}}

\subsection{The Main Sample\label{main}}

We were able to obtain secure spectral classifications for most sources down to an emission-line
surface brightness $> 5.0 \times 10^{-17}{\rm erg\ cm}^{-2}{\rm s}^{-1}{\rm arcsec}^{-2}$. 
We observed 14 candidates down to this limit. 10 were securely classified as $z \sim 2.38$
Ly$\alpha$ emitting galaxies, and two more were marginally confirmed as the same. One of the 
remaining sources was the bizarre B5, discussed below, and the other showed 
clear stellar absorption and is hence a foreground source. 
Three of the securely confirmed sources had previous spectral confirmation \citep{fra97}.

At fainter surface brightnesses our sucess rate was lower. Of the 21 remaining sources we observed,
we were only able to obtain a secure spectral classification for 6, and a marginal classification 
for another 9. All were classified as $z \sim 2.38$ Ly$\alpha$ emitting galaxies.
No spectral features were detected in the remaining 6 sources. 

\begin{figure*}
\plotone{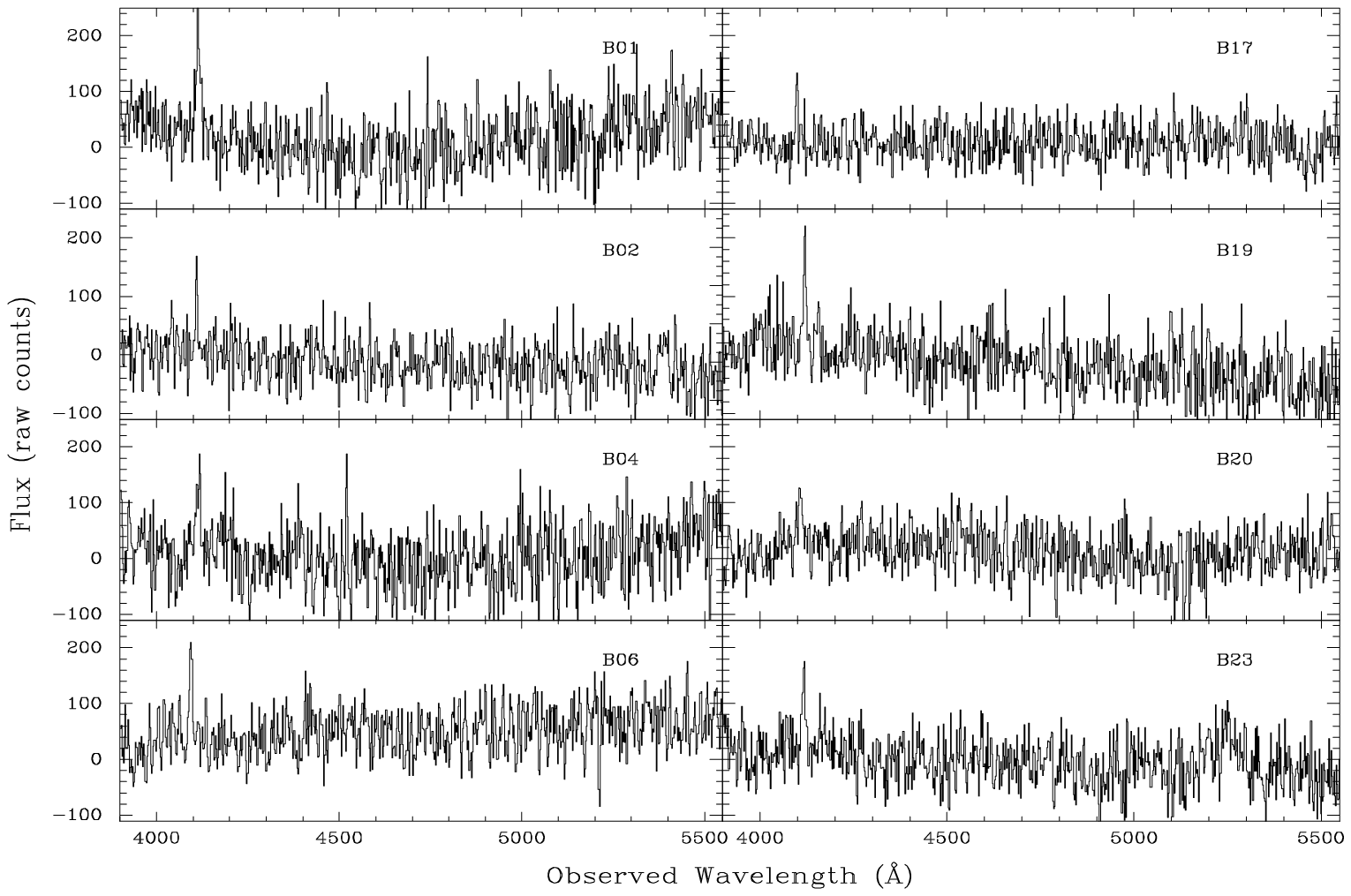}
\plotone{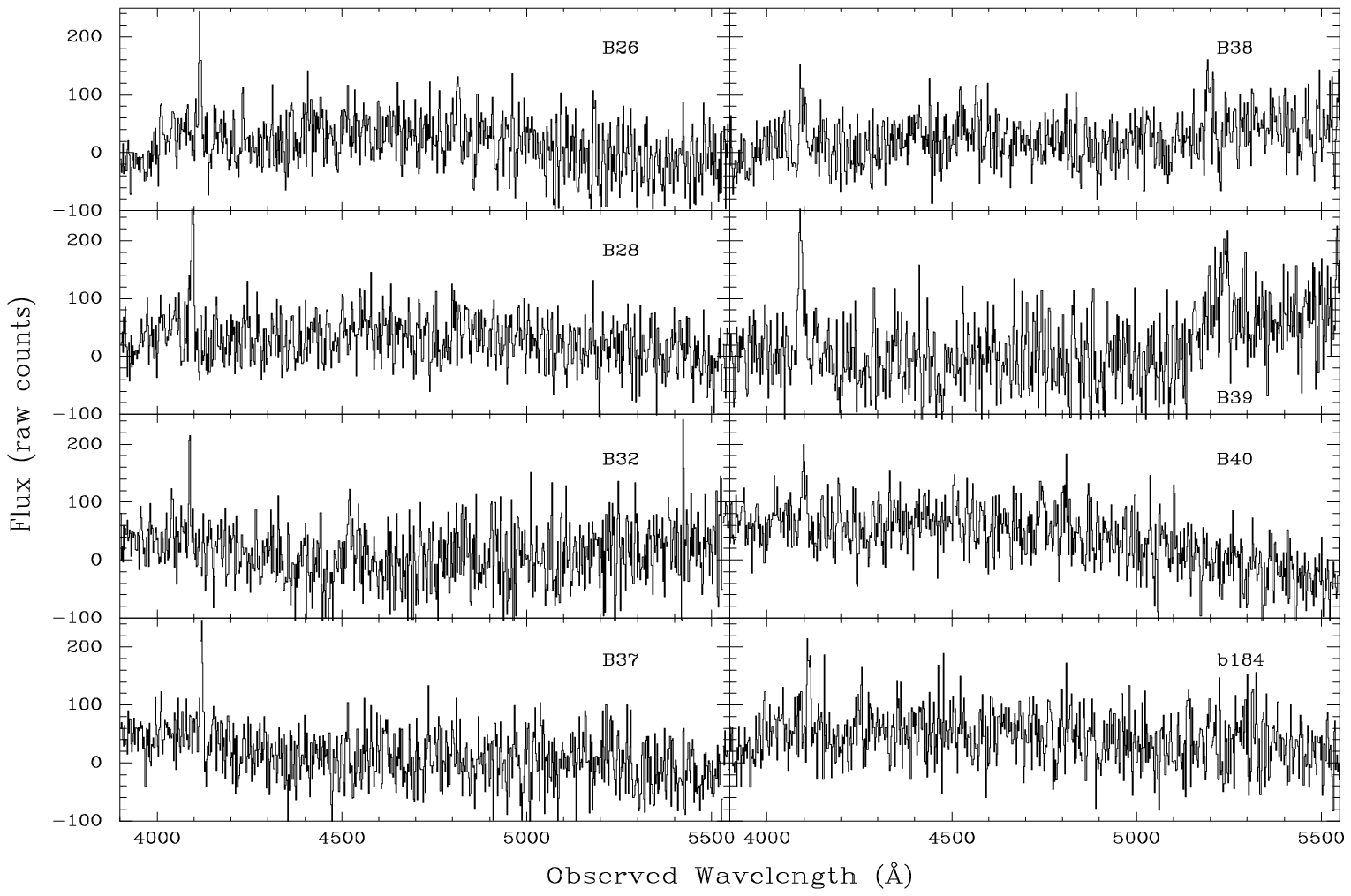}
\caption{
Spectra of confirmed ($> 5 \sigma$) z=2.38 Ly$\alpha$ emitting galaxies. The
spectra are not flux calibrated.
\label{confirmed1}}
\end{figure*}

\begin{figure*}
\plotone{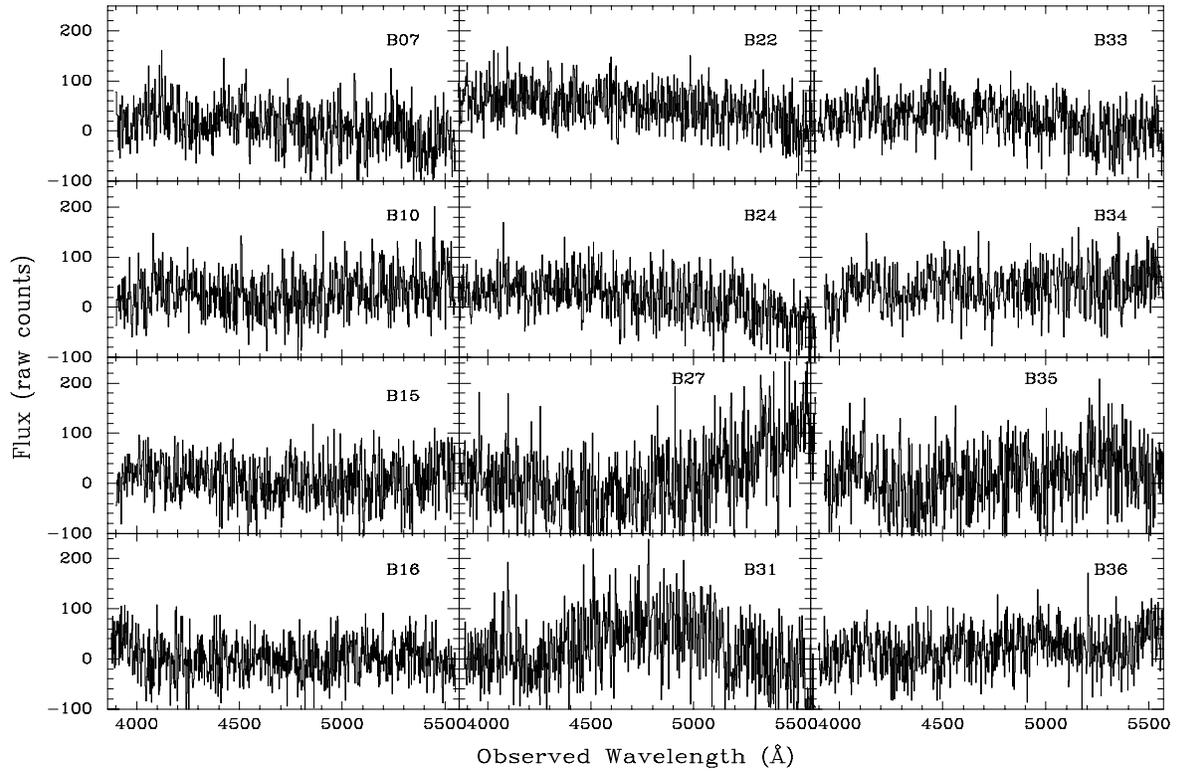}
\caption{
Spectra of the marginal (3 -- 5 $\sigma$) z=2.38 Ly$\alpha$ emitting galaxy
candidates. The spectra are not flux calibrated.
\label{marginalp}}
\end{figure*}

We show spectra of the securely confirmed $z \sim 2.38$ Ly$\alpha$ emitters in Fig~\ref{confirmed1},
and list their properties in Table~\ref{confirmed}.
Those with only marginally (3 -- 5$\sigma$) detected Ly$\alpha$ lines are shown in Fig~\ref{marginalp}
and their properties listed in Table~\ref{marginalt}.

\begin{figure*}
\plotone{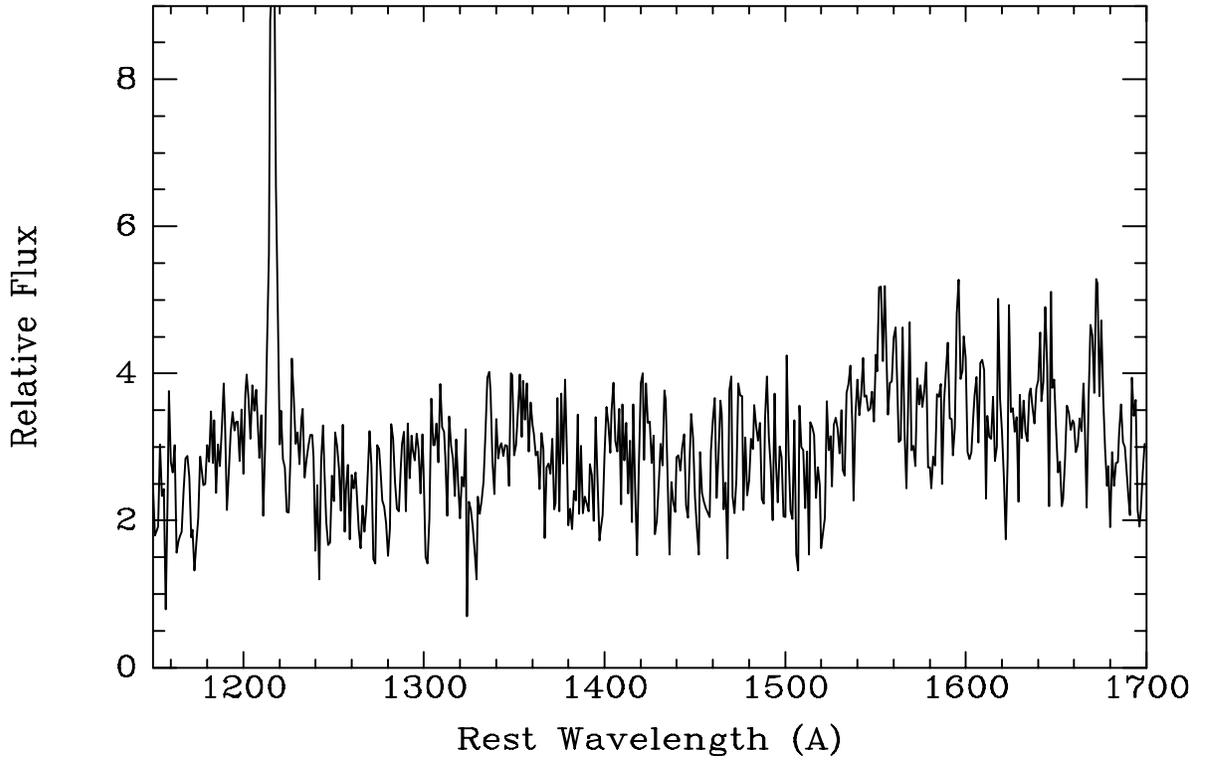}
\caption{
Coadded spectrum of the candidate $z \sim 2.38$ galaxies with a securely 
detected Ly$\alpha$ line.
\label{secure_comp}}
\end{figure*}

\begin{figure*}
\plotone{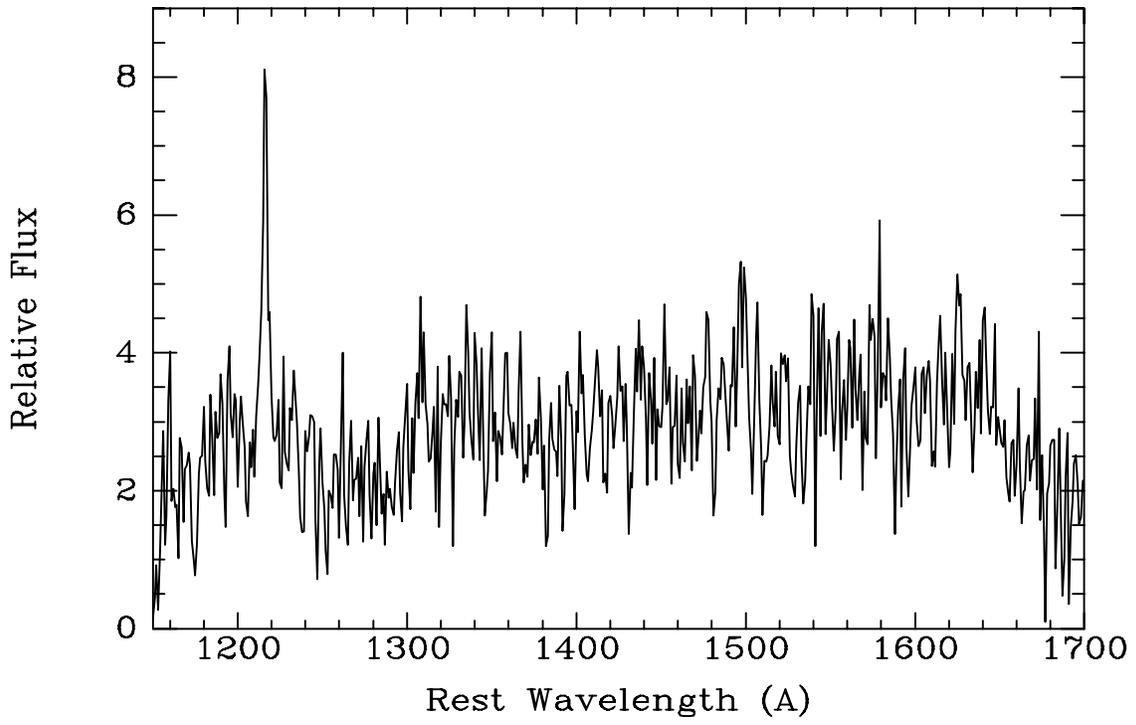}
\caption{
Coadded spectrum of the candidate $z \sim 2.38$ galaxies with a marginally detected Ly$\alpha$ line.
\label{marginal_comp}}
\end{figure*}

Composite spectra of all sources classified as $z \sim 2.38$ Ly$\alpha$ emitting galaxies
were constructed (Figs~\ref{secure_comp} \& \ref{marginal_comp}) by shifting the spectra to 
align the centroid of the putative Ly$\alpha$ line at a nominal wavelength of 1216\AA . If a 
significant fraction of these sources were foreground [\ion{O}{2}] emitting galaxies, we would
expect to see H$\beta$ and [\ion{O}{3}] in these composite spectra, shifted to 1585\AA\ and
1633\AA\ respectively. Nothing is seen at these wavelengths: instead C~IV is weakly detected in the
composite of the securely confirmed sources. 

\begin{figure*}
\plotone{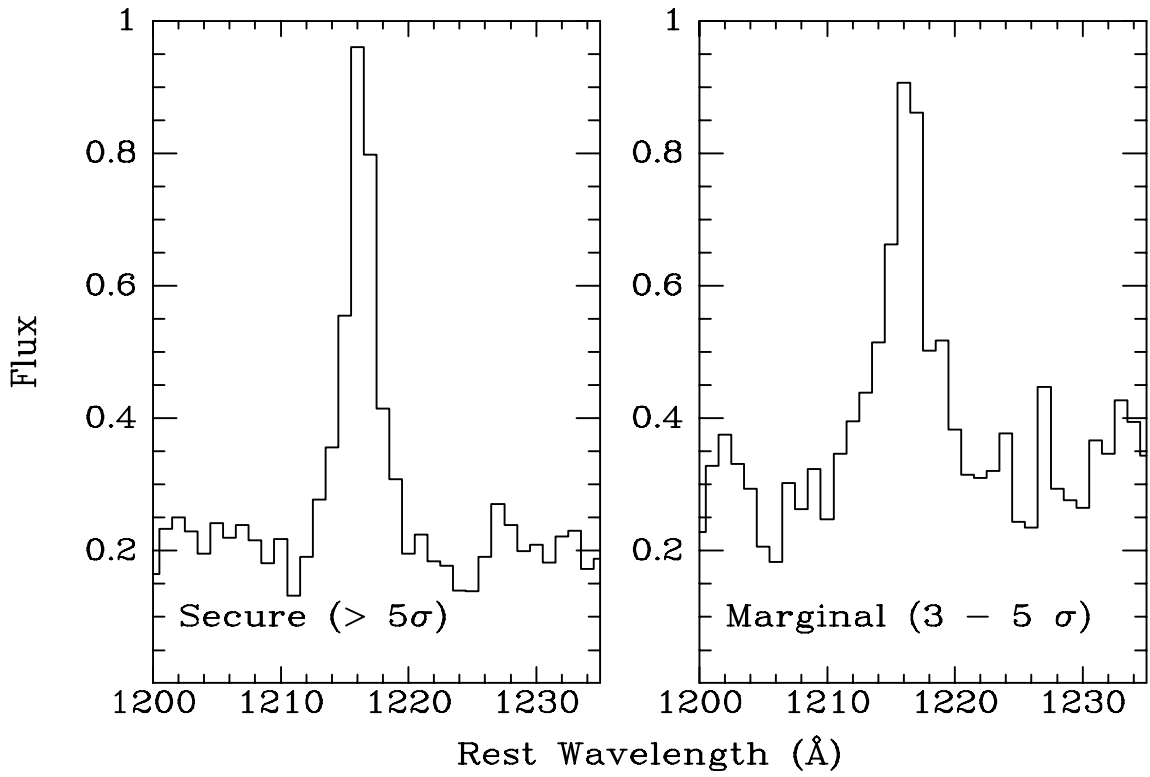}
\caption{
Close-up of the Ly$\alpha$ wavelength region of the two composite spectra in the previous two figures
\label{asym}}
\end{figure*}

A close-up of the Ly$\alpha$ region of these composite spectra
(Fig~\ref{asym}) does not show evidence of asymmetry in the profiles. The individual spectra are
too poor to check for asymmetry, though in our previous spectrum of B1, strong asymmetry was seen
\citep{fra96}. This does not rule out a systematic asymetry in these lines, as it could have 
been washed out by centroiding errors in these noisy spectra. We are unable to look for
evidence of a continuum break across the Ly$\alpha$ line due to the difficulty in accurately
sky-subtracting 2dF spectra. 

\begin{figure}
\plotone{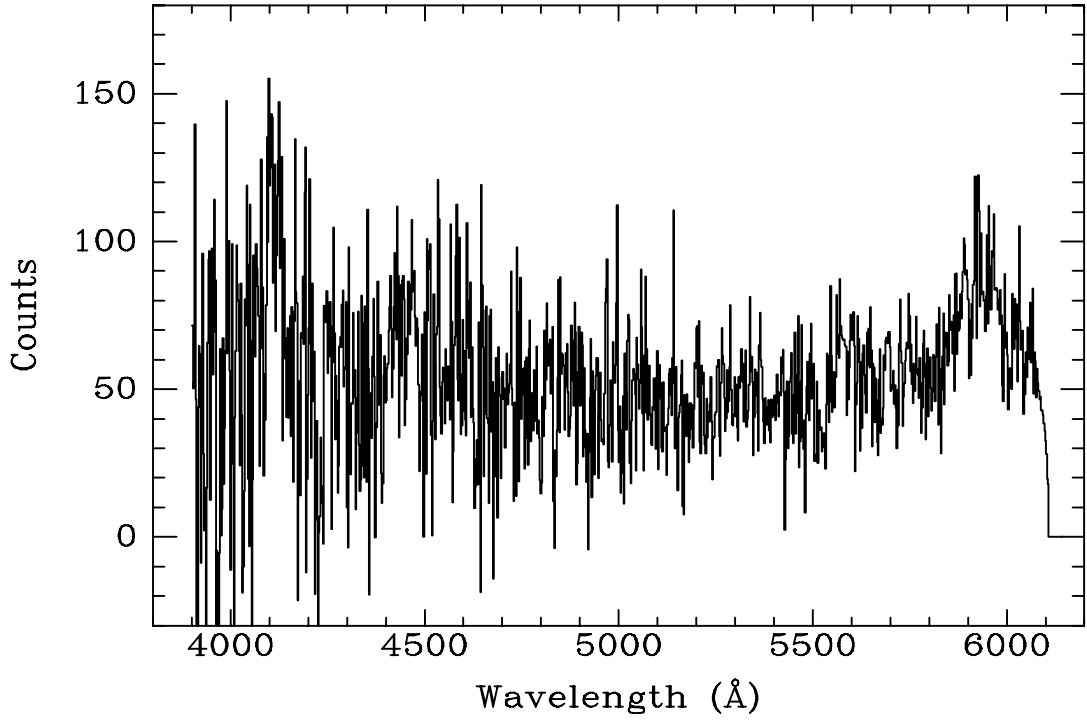}
\caption{
Spectrum of B5. The spectrum is not flux calibrated.
\label{B5spec}}
\end{figure}

One source, B5 (coordinates 21:43:03.57 -44:23:44.2, J2000), proved impossible 
to classify. B5 is one of the most luminous narrow-band excess sources, and
in Paper~1 we classified it as a Ly$\alpha$ blob. Our image of it shows 
that the excess narrow-band flux is centrally concentrated, but clearly extended. 
Our spectrum (Fig~\ref{B5spec}) shows two clear
broad emission lines: one at 4110\AA\ and the other at 5930\AA . Both lines,
though noisy, are independently seen in the co-added data from each night, and
must therefore be regarded as secure. A more marginal candidate broad line is
possibly seen at 4440\AA . The breadth of the lines ($\sim 4000 {\rm km\
s}^{-1}$) clearly indicates that this is a QSO. The fact that B5 shows
extended narrow-band fuzz suggests that the line at 4110\AA\ is either
Ly$\alpha$ or [\ion{O}{2}]. If the former, however,  the 5930\AA\ line would
have a rest-frame wavelength of $\sim 1750$\AA , which does not
correspond to any normal QSO emission line \citep{fra91}. If the latter,
we do not see any Balmer lines or [\ion{O}{3}], and the 5930\AA\ line would lie
at rest-frame 5390\AA ; another wavelength which does not correspond to any
strong QSO line. No alternative identification of the lines works any better.
A higher quality spectrum with wider wavelength coverage is clearly needed.

Three of the securely confirmed $z \sim 2.38$ galaxies show significant
C~IV in emission. Two of these (B38 and B39)
have Ly$\alpha$ velocity widths greater than $1000 {\rm km\ s}^{-1}$ and are thus probable 
QSOs. The remaining source (B23), while its Ly$\alpha$ emission is relatively narrow 
($800 {\rm km\ s}^{-1}$) appears to have much broader  C~IV emission. These sources may thus all
be QSOs, albeit width host galaxy light contributing to their spatially extended images.

Our spectroscopy thus demonstrates that down to a narrow-band flux surface brightness limit of
$5.0 \times 10^{-17}{\rm erg\ cm}^{-2}{\rm s}^{-1}{\rm arcsec}^{-2}$, the vast majority of 
our candidates are indeed Ly$\alpha$ emitting galaxies at $z \sim 2.38$. There is nothing 
in our data to suggest that this fraction is not comparable for the fainter sources, but deeper 
data will be required to confirm this.

\subsection{Other Samples: QSOs and Foreground Galaxies}

Thirteen of the twenty point source sample objects observed were QSOs with
a broad emission line within the narrow-band filter bandpass. Only one of
these, however,  (Fig~\ref{qsos}) was at the filament redshift: ten lie at $z \sim 1.65$ 
which places
\ion{C}{4} within the passband. The other two lie at z=0.457 and z=1.458,
placing \ion{Mg}{2} and \ion{He}{2} respectively in the passband. Two of the 
remaining
sources had no significant features in their spectra, and the other two were
low redshift compact emission-line galaxies (one with [\ion{O}{2}] in the
passband).

\begin{figure*}
\plotone{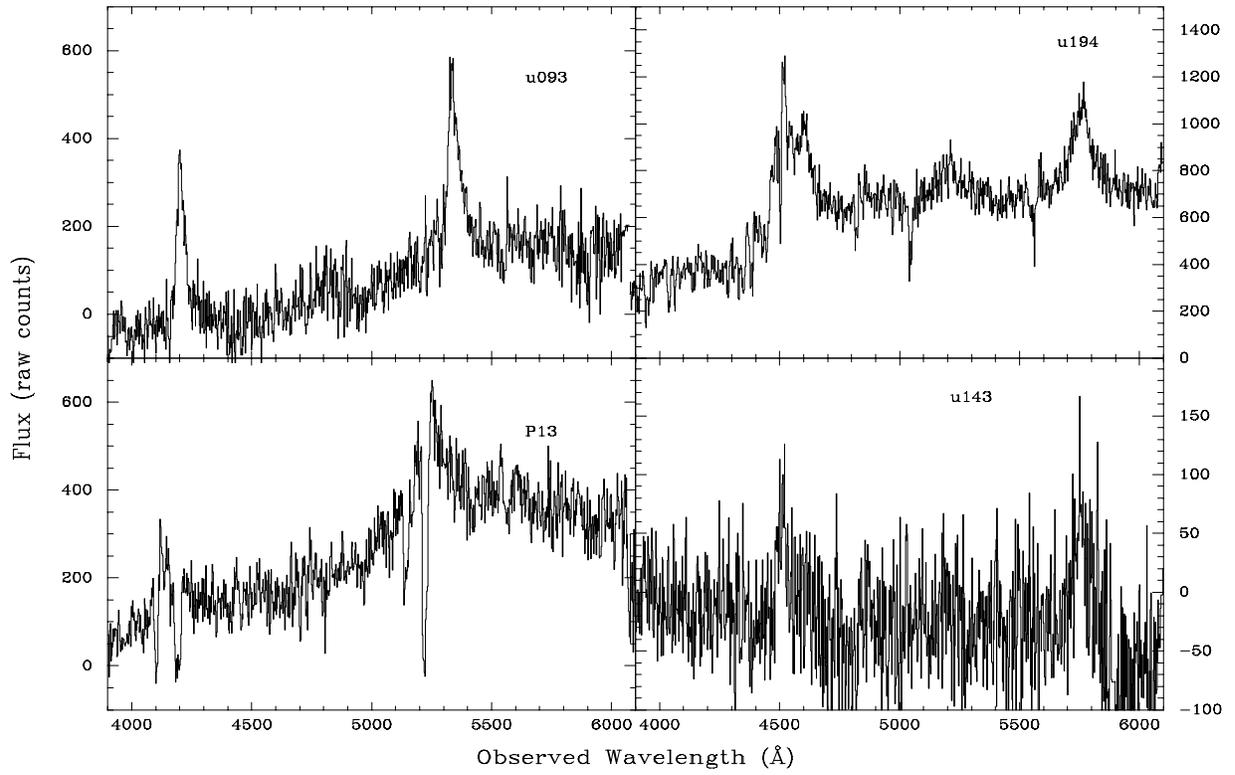}
\caption{
Spectra of QSOs lying at or behind the filament redshift. The spectra are not
flux calibrated.
\label{QSOspec}}
\end{figure*}

A further 41 QSOs were found in the UVX sample, and three in the low
equivalent width sample. Three of the UVX QSOs lie at higher redshifts than
our candidate filament, and are shown in Fig~\ref{QSOspec}. All QSOs
identified are listed in Table~\ref{qsos}. Given our restricted wavelength
coverage, we often see only a single emission line: these one line sources
are noted in the table, and given redshifts typically assuming that this line
is \ion{Mg}{2}. QSOs with lines only marginally detected are indicated with
question marks.

We detect 19 galaxies whose redshifts place [\ion{O}{2}] within the passband of
our filter. These, together with other foreground emission-line
galaxies, are listed in Table~\ref{gals}.

\section{Discussion}

\subsection{Are the Filament and the Voids Real?}

In Paper~1, we showed that the two-dimensional positions of the candidate $z \sim 2.38$ galaxies
on the sky was non-random, and that the two-dimensional void probability function measured from
these candidates was marginally inconsistent with one set of cold dark matter simulations.

We repeated this analysis, adding in the confirmed $z \sim 2.38$ QSO but 
removing the confirmed foreground source. This made no appreciable difference to the results
from Paper~1. We then repeated the analysis using only those candidates with secure
spectral confirmation, and again combining these with the marginally confirmed candidates. Both
analyses gave results consistent with those in Paper~1, but with larger error bars. In neither 
case, however, was the measured void probability function inconsistent with the cold dark matter
simulations with 95\% confidence. The three-dimensional distribution of the galaxies is
discussed in \S~\ref{vel}.

\begin{figure}
\plotone{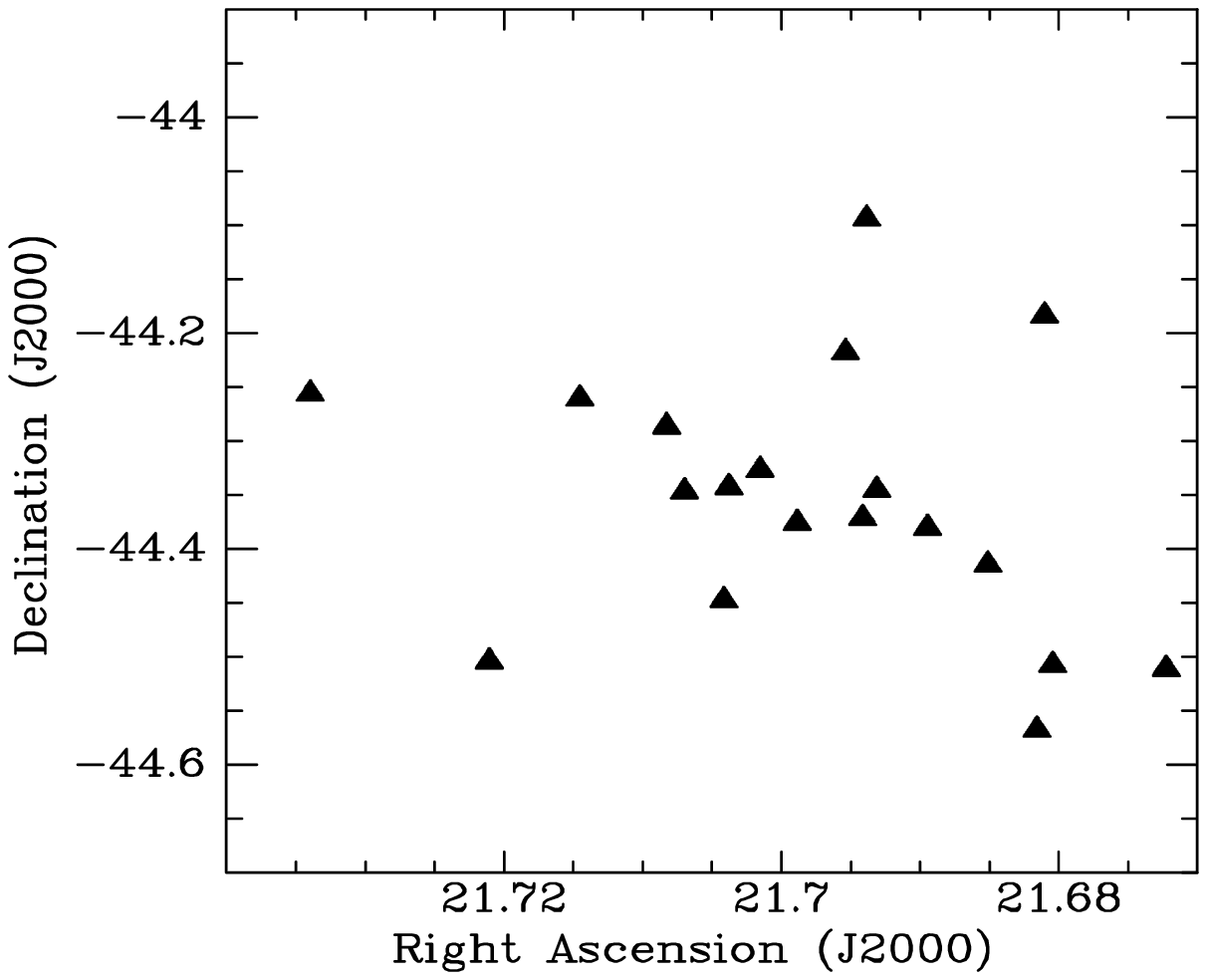}
\caption{
Distribution of foreground $z \sim 0.1$, [\ion{O}{2}] emitting galaxies 
in the field.
\label{o2pos}}
\end{figure}

\begin{figure}
\plotone{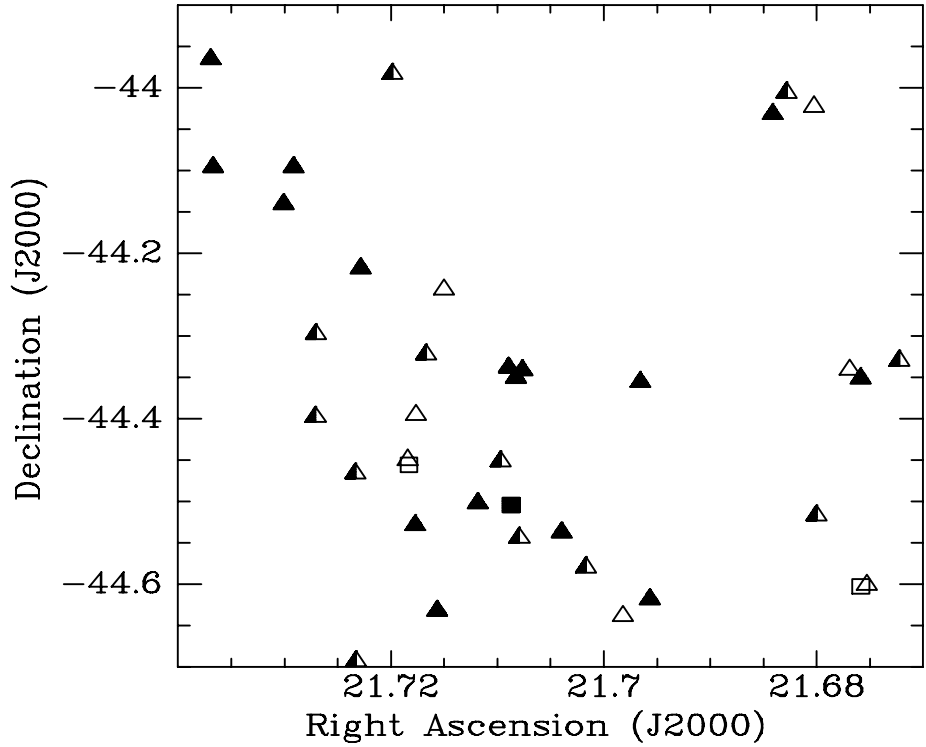}
\caption{
Distribution of candidate Ly$\alpha$ emitting galaxies on the sky. Objects with
secure spectral confirmations are plotted as solid triangles. Marginal
confirmations are shown as half filled triangles, while objects with
featureless spectra are empty triangles. The two unobserved candidates are
shown as empty squares, and the QSO at the filament redshift as a 
filled square.
\label{filplot}}
\end{figure}

We show in \S~\ref{main} that for sources above our surface brightness limit, at least 70\% 
lie at $z \sim 2.38$. We further show that there is no spectral evidence that any of the main
sample sources are foreground [\ion{O}{2}] emitting galaxies at $z \sim 0.1$.
A further test for any population of foreground interlopers is the distribution of our
candidates on the sky. In Fig~\ref{o2pos} and Fig~\ref{filplot}, we show that the
projected distribution of confirmed foreground $z \sim 0.1$ galaxies is quite different from that of the candidate $z \sim 2.38$ galaxies, being concentrated on the cluster Abell 3800 \citep{abe89}. 

Fig~\ref{filplot} also demonstrates that the sources for which we have secure 
spectral confirmations are scattered throughout the cliamed filament, rather than all being
gathered in one region. There is a relative deficit of securely confirmed 
sources in the center of the filament, but the three confirmed sources that do lie in this
region were those studied by \citet{fra97}, all of which have secure multi-line
spectroscopic confirmation. The redshifts of the sources at the top and bottom of the 
filament show no significant differences, so there is no evidence in this data set for 
the filament being simply two clusters with a few stray sources lying between them, though 
this model cannot be ruled out.

\subsection{Is it a Filament or a Sheet?\label{vel}}

If the putative filament were inclined to our line of sight, we might expect to see a 
redshift gradient along it, as was seen by \citet{mol01} in their much smaller filament.
No such correlation was seen: a plot of Ly$\alpha$ wavelength against position along the
filament contains neither a gradient nor any sub-structure.

\begin{figure}
\plotone{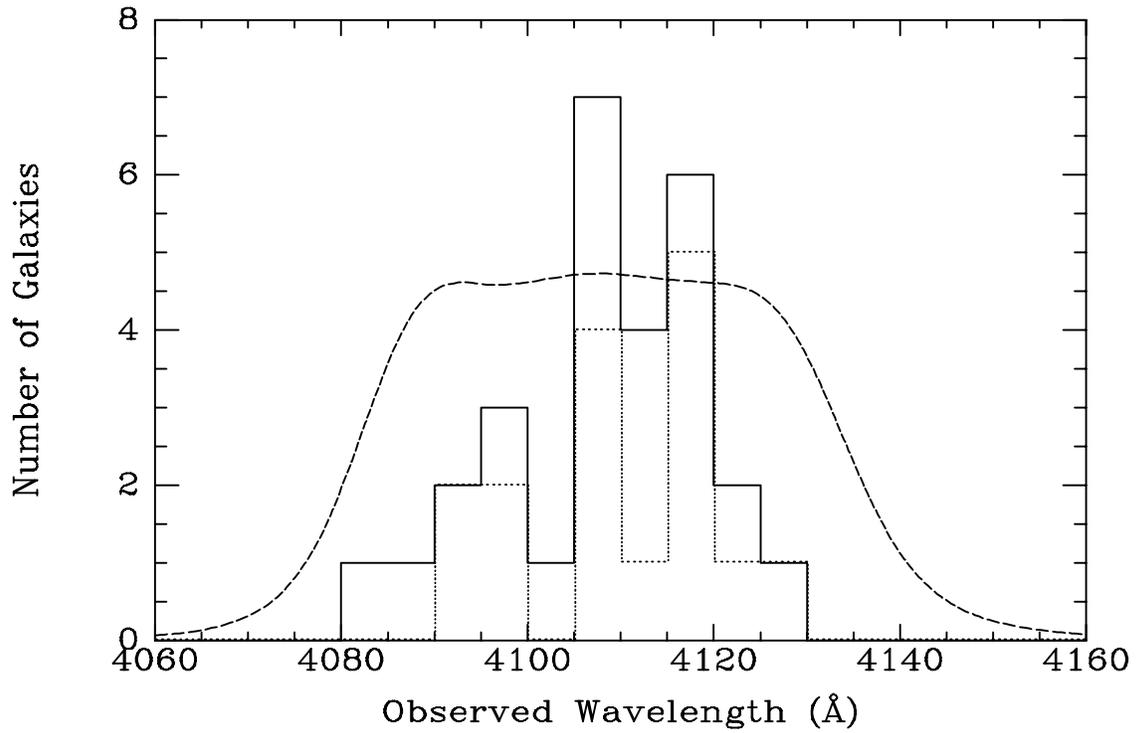}
\caption{
Distribution of wavelengths of the secure and marginal Ly$\alpha$ emitting
galaxies (solid line) and confirmed sources only (dotted line, slightly offset
for clarity), compared to the filter response curve of our narrow-band
filter (dashed line), shifted and broadened as appropriate when placed in
the converging beam at CTIO prime focus.
\label{Lyhist}}
\end{figure}

The distribution of all Ly$\alpha$ redshifts is shown in Fig~\ref{Lyhist}.
The mean redshift (including both secure and marginal candidates) is
$z=2.3791$, with a standard deviation of $820 {\rm km\ s}^{-1}$. The
corresponding figures for sources with secure redshifts only are $z=2.3807$
and $800 {\rm km\ s}^{-1}$.

The standard deviation in the observed redshifts is comparable to the
average Ly$\alpha$ line width of the individual galaxies ($860 {\rm km\ s}^{-1}$, or
 $730 {\rm km\ s}^{-1}$ if we subtract in quadrature the instrumental resolution, 
 Table~\ref{confirmed}).
As the Ly$\alpha$ line is probably extremely optically thick, one would expect the observed 
line peak to be offset from the true systemic redshift by of order the line width. Thus the
scatter in the observed redshifts may be purely an artifact of the optical depths: we have
no significant evidence for an intrinsic scatter. Measurement of the velocity structure will 
require observations in a line less optically thick.

Are we really seeing a concentration of sources at a particular wavelength, or is the
measured wavelength dispersion simply an artifact of the bandpass of our
selection filter? Fig~\ref{Lyhist} suggests that the observed redshift
distribution is more sharply peaked than the filter bandpass, and that the
peak lies at a slightly longer wavelength. We tested this by Monte-Carlo
simulating samples of randomly distributed galaxies observed through this
filter, assuming a Ly$\alpha$ luminosity function with 
$N(L) \propto L^{-0.87}$, as used in \citet{pal04}. Slightly fewer than 
0.1\% (0.6\% for secure sources only) of the simulations have measured 
velocity dispersions
as low as we obtained.

The statistical difference is mostly driven by the relative lack of observed
lines lying in the wavelength range 4085 --- 4105\AA : ie. the observed
galaxy distribution has a lower redshift cut-off. But does it have a higher
redshift cut-off, or is the lack of observed galaxies with Ly$\alpha$
wavelengths longer than around 4125\AA\ caused only by the filter bandpass?
Once again, we tested this using our Monte-Carlo simulations. Slightly fewer 
than
1\% of simulations having the same sample size as the combined secure and
marginal samples had as few sources long-ward of 4125\AA\ as we saw, but if
we restrict ourselves to sources with secure spectral classifications only,
this fraction rises to 10\%.

We conclude that we are probably looking  at a filament viewed from within
$\sim 30^{\circ}$ of sideways,
rather than an edge-on sheet. There remains a small possibility that it is
a sheet but that we are only sampling its front edge, as we do not
definitively detect a higher redshift cut-off.

\subsection{QSO Clusters}

Only one QSO was found in the point source sample at the filament redshift: would 
we have expected more? In the point source sample, we are effectively finding QSOs down to 
$B \sim 21.8$. Extrapolating the QSO
luminosity functions of \citet{boy00}, allowing for QSOs to be selected in
our narrow-band filter even if only part of their broad Ly$\alpha$ emission
is within the filter bandpass and for our incomplete spectroscopy, we would expect 
0.75 QSOs to have been found,
if the region studied was not overdense in them. Thus the observations are
quite consistent with this region being average.

In Paper~1 we reported that the filament region had an overdensity of
Ly$\alpha$ emitting galaxies of $\times 4$. Is this consistent with the
observation of only one QSO? Using Poisson statistics, if the region were
overdense by a factor of 4, we'd find one or fewer QSOs 20\% of the 
time. Thus finding only one QSO is consistent with such an overdensity, 
though favoring lower values.

Perhaps the greatest surprise is the 10 QSOs identified with $1.61 < z < 1.69$,
which places some or all of their \ion{C}{4} emission in the narrow-band
filter. From the \citet{boy00} luminosity functions, we'd predict only 1.8
QSOs in this range. The Poisson probability of finding 10, if the field were
not overdense, is only 0.5\%. We therefore conclude that a cluster of QSOs lies
in the foreground of our field at $z=1.65$. Note that all ten were identified by
their UVX emission as well as their excess narrow-band flux.

\begin{figure}
\plotone{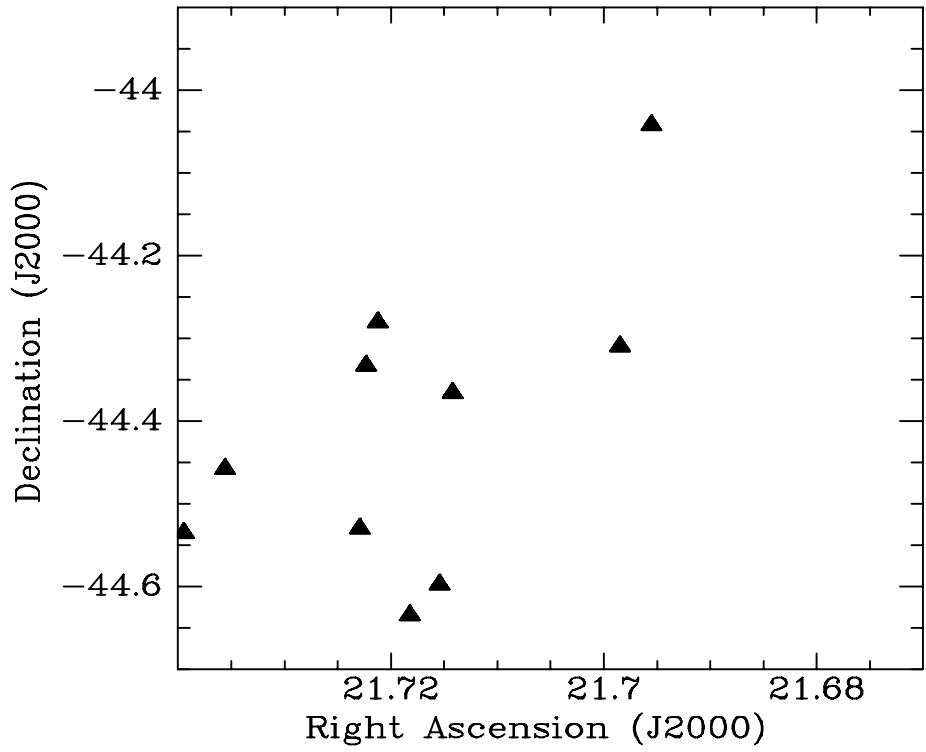}
\caption{
Distribution of foreground $1.61 < z < 1.69$ QSOs (QSOs whose \ion{C}{4} line
lies within the narrow-band filter passband).
\label{QSOcluster}}
\end{figure}

In Fig~\ref{QSOcluster} we show the distribution of these QSOs. It is clearly
non-random: they are concentrated into the South-East corner of the field.
The overdensity of QSOs within this region is $>3$ with 95\% confidence.

Could gravitational lensing by this cluster, plus Abell 3800, be responsible
for the filament and voids seen at z=2.38? Cluster lensing could certainly
be amplifying a handful of individual sources, but expected magnifications
over the enormous extent of our field are insignificant.

\section{Conclusions}

The spectroscopy presented here demonstrates that the brighter candidate filament 
members do indeed lie at redshift $2.38$, and that the candidate list is not extensively
contaminated by $z=0.1$ interlopers. We can also tentatively conclude that the galaxies
we see do indeed lie in a one-dimensional filament, and not
a slice through some two dimensional sheet. These results lend credence to our claim
(in Paper~1) that galaxies at high redshifts are distributed in filaments separated by voids.
A definitive test of this claim will, however, require the systematic mapping of more and
larger volumes of the high redshift universe.

Curiously, the filament region is not overdense in QSOs, but we find a 
significant cluster of foreground QSOs at $z \sim 1.65$. One QSO was found
lying within the filament, as were three background QSOs. Follow-up spectroscopy
of these QSOs may allow us to constrain the gas within the filament and voids,
as has already been done for the three background QSOs previously known
\citep{fra04}.

\acknowledgments

This study was funded by a NASA grant
NRA--98--03--UVG--011, and supported by the STIS IDT through the National 
Optical Astronomical Observatories and by the Goddard Space Flight Center.

\newpage

\newpage
\begin{deluxetable}{lccccc}
%\tabletypesize{\scriptsize}
\tablecaption{Confirmed ($> 5 \sigma$) Ly$\alpha$ Emitting Galaxies at 
z=2.38 \label{confirmed}}
\tablehead{
\colhead{Name} & 
\colhead{Ly$\alpha$} &
\colhead{Position} & 
\colhead{NB} & 
\colhead{B$-$NB} & \colhead{FWHM Ly$\alpha$} \\
\colhead{ ~} & \colhead{Wavelength} & \colhead{(J2000)} & \colhead{(mag)} &
\colhead{(mag)} & \colhead{${\rm km\ s}^{-1}$}
}
\startdata

B1 & 4113  & 21:42:27.56 -44:20:30.1 & 20.97 & 2.47 & 900 \\
B2 & 4109  & 21:42:29.73 -44:21:02.7 & 22.86 & 2.30 & $<$450 \\
B4 & 4117  & 21:42:32.20 -44:20:18.6 & 22.90 & 2.53 & 1300 \\
B6 & 4094  & 21:42:42.63 -44:30:09.0 & 22.13 & 1.89 & 650 \\
B17 & 4097  & 21:41:02.90 -44:01:55.9 & 22.13 & 1.89 & 650 \\
B19 & 4119  & 21:41:44.41 -44:37:06.7 & 22.13 & 1.89 & 730 \\
B20 & 4105  & 21:41:47.66 -44:21:21.9 & 22.61 & 2.37 & 1200 \\
B23 & 4116  & 21:42:14.28 -44:32:15.8 & 22.18 & 2.07 & 700 \\
B26 & 4117  & 21:42:56.34 -44:37:56.8 & 22.48 & 1.56 & $<$450 \\
B28 & 4097  & 21:43:03.80 -44:31:44.9 & 22.16 & 2.30 & 730 \\
B32 & 4090  & 21:43 22.22 -44:13:06.5 & 22.16 & 2.30 & $<$450 \\
B37 & 4122  & 21:43:44.92 -44:05:46.3 & 22.53 & 1.66 & 510 \\
B38 & 4107  & 21:43 48.30 -44:08:26.9 & 22.53 & 1.66 & 1400 \\
B39 & 4106  & 21:44:12.15 -44:05:46.6 & 21.46 & 2.41 & 880 \\
B40 & 4116  & 21:44 12.97 -43:57:56.7 & 21.46 & 2.41 & 1160 \\
b184 & 4129 & 21:44:12.15 -44:05:46.6 & 21 46 & 2.41 & 870 \\

\enddata
\end{deluxetable}

\begin{deluxetable}{lccccc}
%\tabletypesize{\scriptsize}
\tablecaption{Possible Ly$\alpha$ Emitting Galaxies at z=2.38 
(3 -- 5 $\sigma$)
\label{marginalt}}
\tablehead{
\colhead{Name} & 
\colhead{Ly$\alpha$} & 
\colhead{Position} & 
\colhead{NB} & 
\colhead{B$-$NB} \\
\colhead{ ~} & \colhead{Wavelength} & \colhead{(J2000)} & \colhead{(mag)} &
\colhead{(mag)}
}
\startdata

B7 & 4118 & 21 42 34.88 -44 27 06.2 & 21.46 & 2.41 \\
B10 & 4082 & 21:40:19.98 -44:19:48.1 & 22.62 & 1.91 \\
B15 & 4121 & 21 40 48.09 -44 31 01.7 & 22.62 & 1.91 \\
B16 & 4101 & 21 40 58.22 -44 00 22.0 & 22.62 & 1.91 \\
B22 & 4107 & 21:42:06.03 -44:34:47.8 & 23.18 & 2.58 \\
B24 & 4087 & 21 42 28.54 -44 32 38.5 & 23.18 & 2.58 \\
B27 & 4106 & 21:43:00.09 -44:19:21.7 & 22.57 & 1.70 \\
B31 & 4110 & 21:43:11.48 -43:59:01.0 & 23.13 & 2.87 \\
B33 & 4111 & 21:43:23.80 -44:41:36.4 & 22.87 & 1.79 \\
B34 & 4110 & 21:43:24.06 -44:27:59.9 & 22.42 & 1.81 \\
B35 & 4097 & 21:43:37.41 -44:17:53.4 & 23.21 & 2.51 \\
B36 & 4109 & 21:43:37.47 -44:23:52.8 & 22.65 & 2.47 \\

\enddata
\end{deluxetable}

\begin{deluxetable}{cccl}
%\tabletypesize{\scriptsize}
\tablecaption{QSOs \label{qsos}}
\tablehead{
\colhead{Position (J2000)} & 
\colhead{Redshift} &
\colhead{B mag} &
\colhead{Comment} 
}
\startdata
21:40:13.40 -44:09:18.7 & 2.045 & 20.78 & \\
21:40:20.82 -44:32:53.1 & 1.863 & 20.69 & \\
21:40:28.71 -43:57:06.9 & 0.793 & 20.82 &  1 liner \\
21:40:30.12 -44:26:06.3 & 0.608 & 22.37 &  1 liner \\
21:40:34.91 -44:06:48.3 & 0.461 & 23.35 & \\
21:40:39.08 -44:03:07.6 & 2.097 & 20.31 &  \\
21:40:52.48 -44:36:21.5 & 0.457 & 19.96 & \\
21:41:12.32 -44:18:25.7 & 0.692 & 22.48 &  1 liner \\
21:41:13.25 -44:07:38.3 & 0.742 & 21.61 &  1 liner \\
\cutinhead{In the final paper version, the table will be truncated here and
available in electronic format}
21:41:13.67 -43:59:53.9 & 0.688 & 22.90 &  1 liner \\
21:41:15.43 -43:56:00.7 & 2.202 & 21.51 & \\
21:41:18.90 -44:20:42.3 & 0.714 & 22.93 &  1 liner \\
21:41:22.88 -44:06:48.8 & 2.106 & 19.95 & \\
21:41:23.97 -44:33:01.3 & 0.630 & 20.27 &  1 liner \\
21:41:32.62 -44:25:23.4 & 0.988 & 21.35 &  1 liner \\
21:41:43.91 -44:02:33.9 & 1.620 & 20.06 & Foreground cluster\\
21:41:54.51 -44:18:36.2 & 1.647 & 20.14 & Foreground cluster\\
21:41:54.94 -44:42:00.8 & 2.224 & 21.89 &  \\
21:41:55.73 -44:21:37.4 & 0.904 & 20.96 &  1 liner \\
21:41:57.61 -44:19:58.8 & 1.429 & 20.91 &  1 liner \\
21:42:04.05 -44:20:12.5 & 1.039 & 21.28 &  1 liner \\
21:42:04.90 -44:24:57.5 & 1.079 & 19.13 & \\
21:42:07.67 -44:03:10.1 & 1.732 & 19.38 & \\
21:42:18.59 -44:08:20.3 & 1.722 & 22.88 &  ? \\
21:42:18.76 -44:15:48.4 & 1.310 & 22.00 &  ? \\
21:42:20.67 -43:59:20.0 & 2.458 & 21.59 & Background\\
21:42:21.83 -43:58:11.2 & 1.112 & 22.72 &  ? \\
21:42:25.04 -44:40:02.4 & 1.458 & 20.30 & ? \\
21:42:29.03 -43:58:38.6 & 1.853 & 21.96 & \\
21:42:31.33 -44:30:16.8 & 2.388 & 21.24 & Filament member \\
21:42:35.13 -44:32:30.6 & 0.850 & 22.34 & 1 liner \\
21:42:38.71 -44:06:39.0 & 1.736 & 21.10 & \\
21:42:39.38 -44:31:15.2 & 2.036 & 22.79 & \\
21:42:43.51 -44:14:25.5 & 1.589 & 20.14 & \\
21:42:43.93 -44:32:50.8 & 1.849 & 21.44 & \\
21:42:51.15 -44:21:58.9 & 1.683 & 21.32 & Foreground cluster\\
21:42:51.50 -44:30:43.2 & 1.795 & 20.26 & \\
21:42:55.56 -44:35:52.9 & 1.639 & 19.51 & Foreground cluster\\
21:43:05.59 -44:38:06.6 & 1.638 & 20.31 & Foreground cluster\\
21:43:11.46 -44:16:20.2 & 0.664 & 22.76 &  1 liner \\
21:43:12.34 -44:03:35.4 & 0.984 & 20.06 & \\
21:43:16.46 -44:16:51.7 & 1.673 & 21.71 & Foreground cluster\\
21:43:20.40 -44:20:00.5 & 1.619 & 21.05 & Foreground cluster\\
21:43:20.56 -44:04:25.3 & 1.044 & 21.62 &  1 liner \\
21:43:22.55 -44:31:49.4 & 1.631 & 20.86 & Foreground cluster\\
21:43:23.39 -44:35:24.0 & 2.725 & 22.02 & Background \\
21:43:23.56 -43:56:38.6 & 2.223 & 21.82 & \\
21:43:24.21 -44:04:52.0 & 1.045 & 19.44 & \\
21:43:29.62 -44:10:11.9 & 1.768 & 21.71 & \\
21:43:32.94 -43:56:31.3 & 1.557 & 19.65 & \\
21:43:37.66 -44:08:01.5 & 2.166 & 22.02 & \\
21:43:48.73 -44:15:25.1 & 1.305 & 22.36 &  1 liner \\
21:43:51.26 -44:06:09.6 & 1.218 & 22.36 & ? \\
21:44:08.11 -44:27:29.8 & 1.688 & 20.08 & Foreground cluster\\
21:44:17.56 -44:07:02.0 & 2.725 & 21.04 & Background \\
21:44:20.92 -44:23:49.8 & 2.162 & 20.49 & \\
21:44:22.04 -44:32:07.9 & 1.648 & 20.13 & Foreground cluster\\
21:44:29.09 -44:12:03.4 & 2.140 & 22.86 & \\
\enddata
\end{deluxetable}

\begin{deluxetable}{cc}
%\tabletypesize{\scriptsize}
\tablecaption{Foreground Galaxies \label{gals}}
\tablehead{
\colhead{Position (J2000)} & 
\colhead{Redshift}
}
\startdata
21:40:20.15 -44:30:41.9 & 0.101 \\ 
21:40:32.69 -44:32:01.2 & 0.053 \\ 
21:40:42.82 -44:12:14.6 & 0.145 \\ 
21:40:49.58 -44:30:29.2 & 0.102 \\ 
21:40:51.69 -44:11:03.5 & 0.097 \\ 
21:40:58.13 -44:22:37.6 & 0.158 \\ 
21:41:03.10 -44:17:54.0 & 0.349 \\ 
21:41:06.21 -44:12:38.0 & 0.204 \\ 
21:41:06.42 -44:24:54.8 & 0.099 \\ 
21:41:08.66 -44:30:49.1 & 0.357 \\ 
\cutinhead{In the paper version, the table will be truncated here and
available in electronic format}
21:41:10.20 -44:29:21.5 & 0.055 \\ 
21:41:19.34 -44:12:56.5 & 0.062 \\ 
21:41:22.07 -44:22:51.2 & 0.104 \\ 
21:41:34.76 -44:02:12.1 & 0.353 \\ 
21:41:35.24 -44:20:45.5 & 0.097 \\ 
21:41:37.88 -44:05:40.3 & 0.102 \\ 
21:41:37.99 -44:29:43.8 & 0.094 \\ 
21:41:38.90 -44:22:18.8 & 0.103 \\ 
21:41:43.34 -44:13:06.1 & 0.102 \\ 
21:41:48.01 -44:16:15.8 & 0.317 \\ 
21:41:48.28 -44:17:34.1 & 0.328 \\ 
21:41:54.00 -44:14:29.3 & 0.055 \\ 
21:41:55.90 -44:22:34.8 & 0.094 \\ 
21:42:02.79 -44:14:44.2 & 0.311 \\ 
21:42:05.50 -44:19:38.5 & 0.102 \\ 
21:42:12.15 -44:14:53.7 & 0.566 \\ 
21:42:13.63 -44:20:35.2 & 0.099 \\ 
21:42:14.27 -44:29:11.3 & 0.264 \\ 
21:42:14.95 -44:26:53.1 & 0.099 \\ 
21:42:25.12 -44:20:50.3 & 0.098 \\ 
21:42:25.39 -44:10:40.7 & 0.406 \\ 
21:42:29.36 -44:03:27.7 & 0.470 \\ 
21:42:29.82 -44:17:12.8 & 0.099 \\ 
21:42:40.54 -44:22:22.7 & 0.562 \\ 
21:42:40.97 -43:58:15.5 & 0.510 \\ 
21:42:51.39 -43:58:16.6 & 0.137 \\ 
21:42:52.30 -44:15:40.2 & 0.101 \\ 
21:42:56.55 -43:59:23.5 & 0.143 \\ 
21:43:06.10 -43:59:04.2 & 0.462 \\ 
21:43:15.81 -44:30:17.3 & 0.100 \\ 
21:43:27.59 -44:27:52.0 & 0.420 \\
21:43:34.40 -44:42:55.6 & 0.038 \\ 
21:44:02.26 -44:15:24.1 & 0.100 \\ 
\enddata
\end{deluxetable}

\end{document}